\documentclass[floatfix,twocolumn,aps,prd,showpacs,amsmath,amssymb]{revtex4}

\usepackage[utf8]{inputenc}
\usepackage{epsfig}
\usepackage{slashed}
\usepackage{amsmath}
\usepackage{amssymb}

\begin{document}

\title{Bosonization, mass generation, and the pseudo Chern-Simons action }

\author{Gabriel C. Magalhães$^{1}$, Van S. Alves$^{1}$, Leandro O. Nascimento$^{2}$, Eduardo C. Marino$^3$}
\affiliation{$^1$ Faculdade de F\'\i sica, Universidade Federal do Par\'a, Av.~Augusto Correa 01, Bel\'em PA, 66075-110, Brazil  
\\
$^2$Faculdade de Ci\^encias Naturais, Universidade Federal do Par\'a, C.P. 68800-000, Breves, PA,  Brazil
\\
$^3$Instituto de F\'\i sica, Universidade Federal do Rio de Janeiro, C.P. 68528, Rio de Janeiro RJ, 21941-972, Brazil
 }

\date{\today}

\begin{abstract}

We discuss several aspects of a generalization of the Chern-Simons action containing the pseudo-differential operator$\sqrt{-\Box}$, which we shall call pseudo Chern-Simons (PCS). Firstly, we derive the PCS from the bosonization of free massive Dirac particles in (2+1)D in the limit when $m^2\ll p^2$, where $m$ is the fermion mass and $p$ is its momentum. In this regime, the whole bosonized action also has a modified Maxwell term, involving the same pseudo-differential operator. Furthermore, the large-mass $m^2\gg p^2$ regime is also considered. We also investigate the main effects of the PCS term into the Pseudo quantum electrodynamics (PQED), which describes the electromagnetic interactions between charged particles in (2+1)D. We show that the massless gauge field of PQED becomes massive in the presence of a PCS term, without the need of a Higgs mechanism. In the nonrelativistic limit, we show that the static potential has a repulsive term (given by the Coulomb potential) and an attractive part (given by a sum of special functions), whose competition generates bound states of particles with the same charge. Having in mind two-dimensional materials, we also conclude that the presence of a PCS term does not affect the renormalization either of the Fermi velocity and of the band gap in a Dirac-like material.

\end{abstract}


\maketitle

\section{INTRODUCTION}\label{INTRO}

After the experimental realization of graphene \cite{grapexp}, quantum electrodynamics (QED) has been used as an efficient tool for describing the electronic properties in planar materials. In this case, the quasiparticles are described, at low energies, through a massless Dirac field and several applications of this fact have been made  \cite{Bernevig}. For related materials exhibiting a sublattice-symmetry breaking, it is possible to show that other planar materials, such as silicene and the transition metals dichalcogenides (TMDs) \cite{2Dmaterials}, are described by a massive Dirac field, opening possibilities of new applications within a quantum-electrodynamical approach where electronic interactions are naturally included.

PQED \cite{marino} (also referred to as Reduced Quantum Electrodynamics RQED \cite{gorbar,teber2012electromagnetic,kotikov2014two}) is a theory formulated in $(2+1)D$, which describes the electromagnetic interactions of charged particles constrained to move on a plane and it is a very useful tool for calculating either new topological states of matter \cite{PRX2015} or renormalized parameters \cite{vozmediano,luis}. It is unitary \cite{unitarity}, local \cite{do1992canonical}, gauge invariant, and it has been shown to be an example of conformal field theory \cite{conformal}. Several results have been obtained from this model so far \cite{CSBPQED,gfactor,Yukawa}, in particular, the formation of electron-hole bounded states (excitons) in TMDs \cite{TMDPQED}.

In a previous work Ref.~\cite{pqed+cs}, we have shown that the usual Chern-Simons term, when coupled to PQED, provides an effective description of the screening of the dielectric constant in two-dimensional materials. In this case, therefore, there is no mass generation for the gauge field, unlike the well-known topological mass generation that occurs in the Maxwell-Chern-Simons theory \cite{Dunne}. Within the nonperturbative regime, the effects of the Chern-Simons term into PQED has also been
investigated in Refs.~\cite{olivares2020influence,carrington}.

The main reason for the absence of mass generation in Ref.~\cite{pqed+cs} is the canonical dimension of the gauge field in PQED, which implies a dimensionless $\theta$-parameter. On the other hand, an important feature of this gauge field is that it may be obtained from the bosonization of free massless Dirac particles in (2+1)D \cite{marino}. The gauge field obtained from the bosonization of free \textit{massive} Dirac particles, nevertheless, has not been obtained up to now. As we shall conclude later, this gauge field has a PCS term, providing a massive $\theta$-parameter and mass generation. This is similar to what happens in the Maxwell-Chern-Simons theory.

Here, we investigate the bosonization of free massive Dirac particles in (2+1) dimensions, which is a generalization of the result in Ref.~\cite{bosonization}. For massive electrons, we conclude that the bosonized action has the PQED term plus the PCS action, given by 
\begin{equation}
i\theta \epsilon_{\mu\nu\alpha} A^\mu \partial^\nu A^\alpha/\sqrt{-\Box}
\end{equation}
in the small-mass limit $m^2 \ll p^2$. Furthermore, we remark that, in the opposite limit $m^2 \gg p^2$, the bosonized action is equal to the Maxwell-Chern-Simons theory. This closes our results about noninteracting particles. Note that, as pointed out in Ref.~\cite{PLBOzela}, the PCS term coupled to PQED may also be obtained from dual transformations of the Higgs-Chern-Simons action. Therefore, we consider a case of Dirac particles coupled with PQED and the PCS term. From that, we calculate the static potential $V(r)$ for an electron coupled to this gauge field. In this case, it is shown that $V(r)$ is given by a non-symmetric potential around a stationary point $r=r_0$, while for $r\ll r_0$ it is given by the Coulomb potential and $r_0 \propto 1/\theta$. This allow us to discuss the formation of bounded pairs of particles in this model. In the static regime, from the analysis of the renormalization group, we use the perturbative approach for calculating the beta functions of the Fermi velocity and the electron mass. These, however, are shown to be the same as in the case with $\theta=0$.

The outline of this paper is the following: In Sec. II, we investigate the bosonization of massive Dirac particles. In Sec. III, we couple PQED with the nonlocal Chern-Simons action and we analyze the formation of bound states. In Sec. IV, we calculate the screening effect on the gauge-field propagator and its consequence for the static interaction potential. In Sec. V, we calculate the anisotropic electron self-energy in the static regime using the two-component representation for the spinor and obtain the renormalization of the mass and the Fermi velocity. We summarize our results in Sec.~\ref{SCREENING EFFECT ON GAUGE FIELD}. In appendix A, we show some details of the renormalization group. Finally, in appendix B, we calculated the screening effect on the static interaction potential using the RPA approach and adopting the 4x4 representation for the Dirac matrices.

\section{BOSONIZATION OF FREE MASSIVE DIRAC FERMIONS}\label{Bosonization of Massive Dirac Fermions}

We start with the massive Dirac theory in the $D$-dimensional Euclidean space-time, given by
\begin{equation}\label{mdirac}
\begin{split}
\mathcal{L}_{\rm{D}}&=\bar{\psi}\left(i\gamma^{\mu}\partial_{\mu}-m\right)\psi, 
\end{split}
\end{equation} 
where $\psi$ is the Dirac field, $\gamma_\mu$ are the Dirac matrices, and $m$ is the electron bare mass. Therefore, the generating functional of the current-current correlation function reads
\begin{equation}
Z_{\psi}[J]=Z^\psi_0 \int D\bar\psi D\psi \ e^{-S_{{\rm D}}+ \int d^D x\,j_\mu J^\mu }, \label{ZD}
\end{equation}
where $j_\mu=e\bar\psi\gamma_\mu \psi$ is the matter current, $Z^\psi_0$ is a normalization constant, and $J_\mu$ is an external source.  Equation (\ref{ZD}) is quadratic in the Dirac field, and, therefore, using Eq.~(\ref{mdirac}) in Eq.~(\ref{ZD}), we may  solve the integral over $\psi$ to find
\begin{eqnarray}
Z_{\psi}[J]&=&\det\left[1+\frac{e\gamma^\nu J_\nu}{(i\gamma^{\mu}\partial_{\mu}-m)}\right] \nonumber \\
&=&\exp\left\{ {\rm Tr} \ln\left[1+\frac{e\gamma^\nu J_\nu}{(i\gamma^{\mu}\partial_{\mu}-m)}\right]\right\} \nonumber \\
&=& \exp\left\{{\rm Tr}\sum_{N=1}^{\infty} \frac{(-1)^{N-1}}{N} \left[\frac{e\gamma^\nu J_\nu}{(i\gamma^{\mu}\partial_{\mu}-m)}\right]^N\right\}, \label{ZD1}
\end{eqnarray}
where we have used the arbitrary constant $Z^\psi_0$ such that $Z_{\psi}[0]=1$. From Eq.~(\ref{ZD1}), it follows that $Z_{\psi}[J]=Z^{\psi}_2[J] Z^{\psi}_{N>2}[J]$, where $\ln Z^{\psi}_2[J]$ is a quadratic term in the external sources $J_\mu$ and $\ln Z^{\psi}_{N>2}[J]$ are the higher-order polynomials in $J_\mu$. In particular, $Z^{\psi}_2[J]$ reads
\begin{equation}
Z^{\psi}_2[J]=\exp\left\{\frac{1}{2}\int d^D x d^D y J_\mu(x) \Pi^{\mu\nu}(x-y) J_\nu(y)\right\}, \label{ZD2}
\end{equation}
where $\Pi_{\mu\nu}$ is the vacuum polarization tensor. It is clear, from these results, that now we may calculate the $N$-point current-current correlation functions. As an example, let us consider the two-point current-current correlation, namely, $\langle j_\mu j_\nu \rangle$, given by
\begin{equation}
\langle j_\mu(x) j_\nu(y) \rangle=\frac{\delta^2 Z_\psi[J]}{\delta J^\mu(x) \delta J^\nu(y)}{\huge{|_{J=0}}}=\Pi_{\mu\nu}(x-y).
\end{equation}

Since we are considering free particles, let us take only the quadratic terms in the external sources, such that $Z_\psi[J]=Z^\psi_2[J]$ properly yields our two-point correlation function. Next, we follow the bosonization method, which has been recently reviewed in Ref.~\cite{marinolivro} and applied for massless Dirac particles in Ref.~\cite{bosonization}. Here, we generalize this method for the massive case. The main goal is to calculate the bosonic version of the action in Eq.~(\ref{mdirac}).

\subsection{Two-Dimensional Electrons}

For two-dimensional electrons, the natural bosonic action is defined by a gauge field $B_\mu$.  The action of this model reads 
\begin{equation}
\mathcal{L}_{\rm{B}}=\frac{1}{2} B_\mu A^{\mu\nu} B_\nu+{\rm GF}, \label{actionh}
\end{equation}
and the generating functional is given by
\begin{equation}
Z_B[J]=Z^B_0 \int DB_\mu \ e^{-S_{{\rm B}}+ \int d^D x\,J_\mu K^{\mu\nu} B_\nu }, \label{ZDH}
\end{equation}
where $A_{\mu\nu}$ and $K_{\mu\nu}$ are unknown tensors and ${\rm GF}$ stands for the usual gauge-fixing term.

Following similar steps, after integrating over $B_\mu$ in Eq.~(\ref{ZDH}), one obtains
\begin{equation}\label{Zh}
Z_B[J]=\exp\left\{\frac{1}{2}\int d^D x d^Dy J^\mu [K_{\mu\sigma} (A^{-1})^{\sigma \lambda} K_{\lambda \nu}] J^\nu\right\},
\end{equation} 
where $Z^B_0$ is chosen such that $Z_B[0]=1$. Note that the gauge-fixing term does not appear in Eq.~(\ref{Zh}) because of current conservation. The bosonization follows from assuming that $Z^\psi_2[J]=Z_B[J]$. Hence, after comparing Eq.~(\ref{ZD2}) and Eq.~(\ref{Zh}), we find that $[K_{\mu\sigma} (A^{-1})^{\sigma \lambda} K_{\lambda \nu}]=\Pi_{\mu\nu}$, which is satisfied by taking the simplest solution $A_{\mu\nu}=K_{\mu\nu}=\Pi_{\mu\nu}$. Using the dimensional regularization scheme for $2\times 2$ Dirac matrices, we find \cite{rao1986parity}
\begin{equation}\label{oneloop}
\Pi^{\mu\nu}(p,m)=\Pi_{1}(p,m)P^{\mu\nu}+\Pi_2(p,m)\epsilon^{\mu\alpha\nu}p_\alpha,
\end{equation}
where
\begin{equation}\label{pi1} 
\Pi_{1}=-\frac{e^2}{2\pi}\int_{0}^{1}dx\frac{p^{2}x(1-x)}{\sqrt{x(1-x)p^{2}+m^{2}}}
\end{equation}
and
\begin{equation}\label{pi2}
\Pi_2=\frac{e^2}{4\pi}\int_{0}^{1}dx\frac{m}{\sqrt{x(1-x)p^{2}+m^{2}}}. 
\end{equation}

Finally, after comparing Eq.~(\ref{ZD}) and Eq.~(\ref{ZDH}), we find
\begin{equation}
\bar{\psi}\left(i\gamma^{\mu}\partial_{\mu}-m\right)\psi = \frac{1}{2} B_\mu \Pi^{\mu\nu} B_\nu \label{kinB1}
\end{equation}
and
\begin{equation}
e\bar\psi\gamma^\mu \psi=\Pi^{\mu\nu} B_\nu. \label{curB1}
\end{equation}
Note that Eq.~(\ref{kinB1}) yields the bosonization relation between the fermionic and bosonic kinetic terms, while Eq.~(\ref{curB1}) gives the fermionic matter current in terms of the bosonic gauge field $B_\mu$, identically conserved in Eq.~(\ref{curB1}). Having this in mind, one may calculate the two-point current-current correlation function as 
\begin{equation}\label{endB}
\langle j^\mu j^\nu \rangle=\Pi^{\mu\alpha}\Pi^{\nu\beta}\langle B_\alpha B_\beta \rangle=\Pi^{\mu\nu}.
\end{equation}
Note that, for deriving the last identity in Eq.~(\ref{endB}), we use the fact that the $B_\mu$-field propagator is the inverse of $\Pi^{\mu\nu}$, as we can infer from the rhs of Eq.~(\ref{kinB1}). As expected, we conclude that the bosonized model yields the same result of the fermionic theory.

Using Eq.~(\ref{oneloop}) in Eqs.~(\ref{kinB1}) and (\ref{curB1}), one finds a general bosonized theory for any value of $m$. Next, for the sake of simplicity, we consider the solution either in the small-mass limit $m^2\ll p^2$ or in the large-mass limit $m^2\gg p^2$. 

\subsubsection{The Case $m^2\ll p^2$}

In this case, the lowest-order terms of Eqs.~(\ref{pi1}) and (\ref{pi2}) are 
\begin{equation}
\Pi_1\approx -\frac{e^2\sqrt{p^2}}{16}\left[1+O\left(\frac{m^2}{p^2}\right)\right] \label{pi1ms}
\end{equation}
and
\begin{equation}
\Pi_2 \approx e^2\frac{m}{4\sqrt{p^2}}\left[1+O\left(\frac{m^2}{p^2}\right)\right], \label{pi2ms}
\end{equation}
respectively. The main step is to replace the Fourier transform of Eqs.~(\ref{pi1ms}) and (\ref{pi2ms}) in Eq.~(\ref{actionh}) with $A_{\mu\nu}=\Pi_{\mu\nu}$. After scaling the gauge-field as $B_\mu\rightarrow \sqrt{32} \tilde{B}_\mu /e$, we find
\begin{equation}
\mathcal{L}_{\rm{NCS}}=\frac{1}{2} \frac{\tilde{B}^{\mu\nu} \tilde{B}_{\mu\nu}}{\sqrt{-\Box}}+\frac{4im}{\sqrt{-\Box}} \epsilon^{\mu\nu\alpha}\tilde{B}_\mu \partial_\nu \tilde{B}_\alpha+{\rm GF}, \label{actionh3}
\end{equation}
where $\tilde{B}_{\mu\nu}=\partial_\mu \tilde{B}_\nu-\partial_\nu \tilde{B}_\mu$ is the field intensity tensor of $\tilde{B}_\mu$ with $[\tilde{B}_\mu]=1$, i.e., the bosonic field has dimension of mass.  Therefore, we conclude that
\begin{eqnarray}
{\cal L}_D&=&\bar{\psi}\left(i\gamma^{\mu}\partial_{\mu}-m\right)\psi  \nonumber \\ 
&\approx &\frac{1}{2} \frac{\tilde{B}^{\mu\nu} \tilde{B}_{\mu\nu}}{\sqrt{-\Box}}+\frac{4im}{\sqrt{-\Box}} \epsilon^{\mu\nu\alpha}\tilde{B}_\mu \partial_\nu \tilde{B}_\alpha, \label{R10}
\end{eqnarray}
which gives the bosonization of the kinetic term and
\begin{eqnarray}\label{R11}
j^\mu &=&e\bar\psi\gamma^\mu \psi  \nonumber \\ 
&\approx &\frac{e}{4\sqrt{2}} \frac{\partial_\nu \tilde{B}^{\mu\nu}}{\sqrt{-\Box}}+iem\epsilon^{\mu\nu\alpha}\sqrt{\frac{2}{(-\Box)}}\partial_\nu \tilde{B}_\alpha,
\end{eqnarray}
providing the bosonization of the matter current. Surprisingly, a new kind of Chern-Simons term naturally appears in Eq.~(\ref{actionh3}). This has a pseudo-differential operator $(-\Box)^{-1/2}$ that resembles the PQED model. On the other hand, this new term has the same classical symmetries as the standard Chern-Simons term, i.e., it is gauge invariant and breaks the parity symmetry. A final remark about classical properties is that the pole of the gauge-field propagator (similarly to the fermionic field) occurs at $p^2_{{\rm Mik}}=m^2$, where ${\rm Mik}$ refers to the Minkowski space. Finally, as expected, note that the matter current in Eq.~(\ref{R11}) is identically conserved. Next, we discuss the opposite limit when $m^2 \gg p^2$. 

\subsubsection{The Case $m^2\gg p^2$}

After considering the limit $m^2 \gg p^2$ in Eq.~(\ref{oneloop}), the lowest-order terms read
\begin{equation}
\Pi_1 \approx \frac{e^2p^2}{12 m \pi} +O\left(\frac{p^3}{m^3}\right)\label{AB1}
\end{equation}
and
\begin{equation}
\Pi_2 \approx \frac{e^2}{4\pi}+O\left(\frac{p^2}{m^2}\right). \label{AB2}
\end{equation}
Similarly to the previous case, we use Eqs.~(\ref{AB1}) and (\ref{AB2}) in Eq.~(\ref{actionh}). After scaling the gauge-field field as $B_\mu \rightarrow \sqrt{12 m \pi} \bar{B}_\mu /e$, we find
\begin{equation}
\mathcal{L}_{\rm{MCS}}=\frac{1}{4} \bar{B}^{\mu\nu} \bar{B}_{\mu\nu}+\frac{3m}{2} \epsilon^{\mu\nu\alpha}\bar{B}_\mu \partial_\nu \bar{B}_\alpha+{\rm GF}, \label{AB3}
\end{equation}
where $\bar{B}_{\mu\nu}=\partial_\mu \bar{B}_\nu-\partial_\nu \bar{B}_\mu$ with $[\bar{B}_\mu]=1/2$. Therefore, we conclude that
\begin{eqnarray}
{\cal L}_D&=&\bar{\psi}\left(i\gamma^{\mu}\partial_{\mu}-m\right)\psi  \nonumber \\ 
&\approx &\frac{1}{4} \bar{B}^{\mu\nu} \bar{B}_{\mu\nu}+\frac{3m}{2} \epsilon^{\mu\nu\alpha}\bar{B}_\mu \partial_\nu \bar{B}_\alpha, \label{R12}
\end{eqnarray}
which gives the bosonization of the kinetic term and
\begin{eqnarray}
j^\mu &=&e\bar\psi\gamma^\mu \psi  \nonumber \\ 
&\approx &\frac{e}{2\sqrt{3m\pi}} \partial_\nu \bar{B}^{\mu\nu}+\frac{ie}{2}\sqrt{\frac{3m}{\pi}}\epsilon^{\mu\nu\alpha}\partial_\nu \bar{B}_\alpha, \label{R13}
\end{eqnarray}
providing the bosonization of the matter current.

Note that Eq.~(\ref{AB3}) is, essentially, the Maxwell-Chern-Simons model, where the Chern-Simons parameter $3m/2$ provides a topological mass for the gauge field. This term breaks the parity symmetry, however, this is not an anomaly because we start with a massive electron. For the sake of completeness, note that when starting with $m\rightarrow 0$, one finds the so-called parity anomaly and the gauge field term is given by PQED, as it has been discussed in Ref.~\cite{bosonization}.

\section{THE PSEUDO CHERN-SIMONS MODEL}\label{THE NONLOCAL CHERN-SIMONS MODEL}

In the previous section, we showed that the bosonization approach of the massive Dirac theory, in the regime where $m^2\ll p^2$, generates an effective Lagrangian given by Eq.~(\ref{actionh3}), in which both Maxwell and Chern-Simons terms are modified by the $(-\Box)^{-1/2}$ factor, and as a consequence of this, the gauge field acquires mass. Here, we consider a model describing the interaction of this gauge field with Dirac particles, given by (in Euclidian space-time),

\begin{equation}\label{modelnm}
	\begin{split}
		\mathcal{L}&=\dfrac{1}{2}\dfrac{F^{\mu\nu}F_{\mu\nu}}{\sqrt{(-\Box)}}+\frac{\lambda}{2}\frac{(\partial^{\mu}A_{\mu})^{2}}{\sqrt{(-\Box)}}+\dfrac{i\theta}{2}\dfrac{\epsilon_{\mu\nu\gamma}A^{\gamma}\partial^{\nu} A^{\mu}}{\sqrt{(-\Box)}}+\\
		&+\sum_{j=1}^{N_{f}}\bar{\psi}_{j}\left(i\gamma^{\mu}\partial_{\mu}-m+e\gamma^{\mu}A_{\mu}\right)\psi_{j}, 
	\end{split}
\end{equation} 
where $N_{f}$ is the number of flavors, $A^\mu$ is the gauge field, $e$ is the dimensionless coupling constant, and $\theta$ is the Chern-Simon parameter. Note that their dimensions, in the natural system of units, are given by $[A]=1$, $[\theta]=1$, and $[e]=0$, respectively.

In general grounds, the main purpose of Eq.~(\ref{modelnm}) is to generalize the PQED model by introducing a massive parameter $\theta$ for the gauge field. In order to do so, and having in mind the canonical dimension of the gauge field $[A_\mu]=1$, it is not difficult to conclude that we must have the pseudo-differential operator $(-\Box)^{-1/2}$ in the Chern-Simons term in Eq.~(\ref{modelnm}).

\subsection{Feynman Rules}\label{Feynman Rules}

The gauge-field propagator of the model in Eq.~(\ref{modelnm}) reads
\begin{equation}\label{propagadordofoton}
\begin{split}
\Delta_{\mu\nu}^{(0)} &=\dfrac{1}{2\varepsilon\sqrt{p^{2}}}\dfrac{1}{(p^{2}+\theta^{2})}\left[p^{2}P_{\mu\nu}+\theta\epsilon_{\mu\nu\alpha}k^{\alpha}\right]+\Delta^{(\textrm{GF})}_{\mu\nu},  
\end{split}
\end{equation}
where $P_{\mu\nu}=\delta_{\mu\nu}-p_{\mu}p_{\nu}/p^{2}$, $p^{2}=p_{0}^{2}+\textbf{p}^{2}$, and the parameter $\varepsilon$ is included in order to describe the dielectric constant. In the limit $\theta\rightarrow 0$, Eq.~(\ref{propagadordofoton}) provides the usual PQED propagator. The gauge-fixing term is given by	
\begin{equation}\label{propgeralGF}
\begin{split}
\Delta_{\mu\nu}^{\textrm{(GF)}}(\lambda,p)&=\frac{1}{2\lambda}\left[\frac{p_\mu p_\nu/p^2}{(p^2)^{1/2}+\theta^2(p^2)^{-1/2}}+\right.\\
&\left.+\frac{\theta^2 p_\mu p_\nu/p^2}{(p^2)^{3/2}+\theta^2(p^2)^{-3/2}}\right], 
\end{split}
\end{equation} 
which vanishes in the Landau gauge $\lambda=\infty$. The Dirac field propagator is
\begin{equation}\label{fermionprop}
S^{(0)}_{F}=\frac{-1}{\gamma^\mu p_\mu-m}.
\end{equation}

The pole of Eq.~(\ref{propagadordofoton}) yields a physical mass at $p^2_M=-p^2=\theta^2$ for the gauge field, where $p^2_M$ is the four-momentum in the Minkowski space. Note that the usual Chern-Simons term generates this mass for the Maxwell field, but it is dimensionless for PQED \cite{pqed+cs}. Therefore, in order to find a mass for the gauge-field in PQED, we need our pseudo Chern-Simons term. In particular, this term preserves the gauge symmetry and breaks parity, which indicates that such description (using PQED) is suitable for describing topological effects instead of spontaneous symmetry breaking. 

Interesting, as shown in Ref.\cite{PLBOzela}, the same effective Lagrangian is obtained through the dualization procedure of the Abelian Chern-Simons-Higgs model. In fact, the dual transformation of ${\cal L}=- i\frac{\Theta}{4}{}\epsilon_{\mu\nu\gamma}A_{\mu}{}F_{\nu\gamma}+|D_\mu \phi|^2 + V(|\phi|)$, where $D_\mu \equiv \partial_\mu + ieA_\mu$ is the covariant derivative, $\Theta$ is the Chern-Simons parameter, and $V(|\phi|)$ is a spontaneous symmetry breaking potential, yields a pseudo Chern-Simons action $\sim -i{\theta} \epsilon_{\mu \nu \gamma} a_\mu\partial_\nu\left[ \frac{1}{(-\Box)^{1/2}}\right] a_\gamma$, where $\theta = 2e^2 \rho_0^2/\Theta$, that plays the role of the new Chern-Simons parameter in dual theory, and $\rho_{0}=|\langle \phi\rangle|$ is the vacuum expectation value of the Higgs field $\phi$. Although the dual model originally describes the interaction between vortices, we can consider other types of matter current, such as the fermionic current coupled to this gauge field. This may have some relevance for the description of two-dimensional materials such as graphene and transition metal dichalcogenides (TMD's) \cite{choi2017recent,TMDPQED}. This approach results in a theory similar to Eq.~(\ref{actionh3}) with the PCS term, but more general, because it is valid for any regime of $m$, independent on the Dirac matrices adopted.

Next, we investigate the main effect of $\theta$ for the matter field in the static limit.

\section{SCREENING EFFECT ON THE GAUGE FIELD}\label{SCREENING EFFECT ON GAUGE FIELD}

\begin{figure}[hbtp]
\centering
\includegraphics[scale=.69]{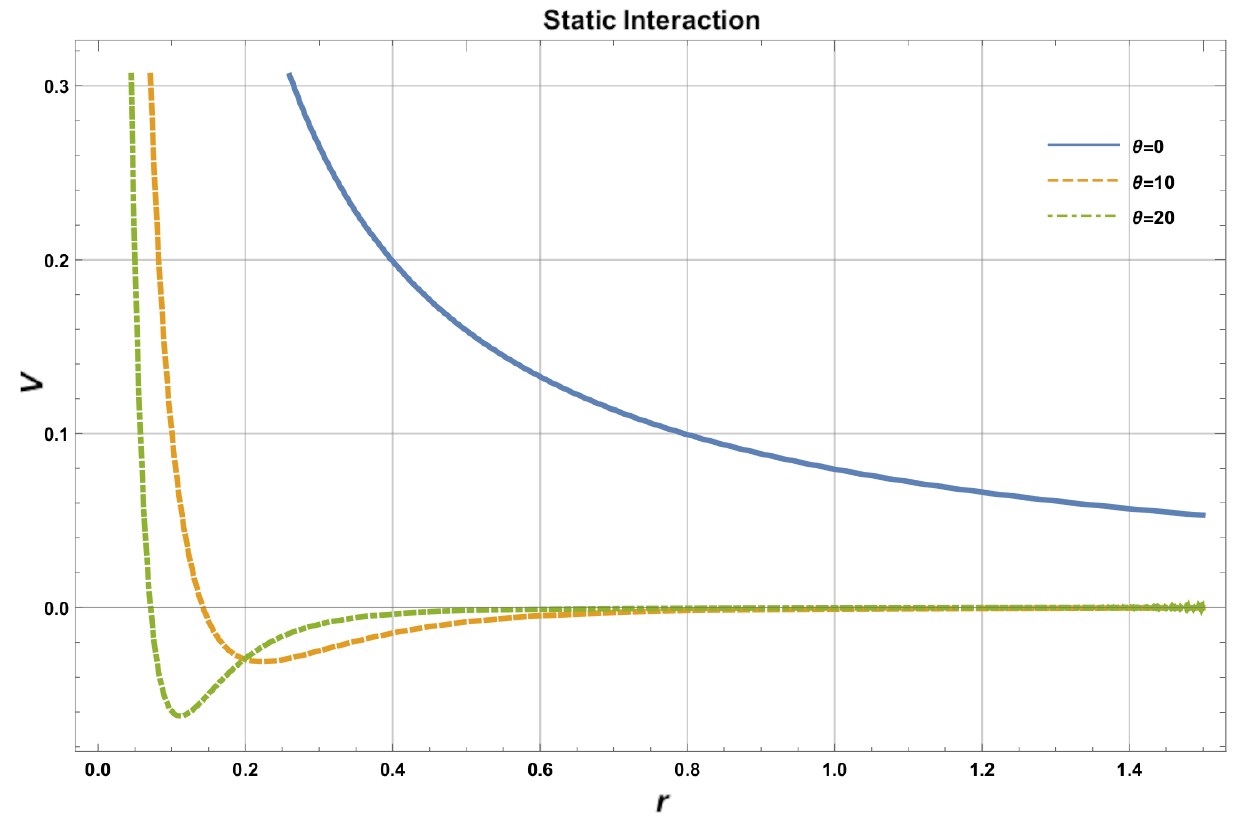}
\caption{The static potential modified by the nonlocal Chern-Simons term. This plot shows the behavior of the static potential in Eq.~(\ref{potencial4}) as a function of the distance $r$ between the fermions, varying the parameter $\theta$. The solid line (blue), dashed (orange), and dotted-dashed (green) show the static potential for $\theta=0$, $\theta=10$, and $\theta=20$ in units of inverse of $r$, respectively.}\label{fig1}
\end{figure}

As shown in Ref.~\cite{PLBOzela}, the static potential for the theory represented by the Lagrangian Eq.~(\ref{modelnm}) is given by
\begin{equation}\label{potencial4}
V(r)=\dfrac{e^{2}}{4\pi\varepsilon}\left\lbrace\dfrac{1}{r}+\frac{\theta\pi}{2}\left[ L_{0}\left(\theta r\right)-I_{0}\left(\theta r\right)\right]\right\rbrace,
\end{equation}
a non-symmetric potential around a stationary point $r_{0}\simeq 2.229/\theta$, where $L_{0}$ is a zero-order Struve L function and $I_{0}$ is a zero-order Bessel I function. Note that, in the limit $\theta\rightarrow 0$, the static interaction is the Coulomb potential, which is an expected feature of PQED. Furthermore, the depth of the potential is given by $V(r_{0})\simeq (-3.0\times 10^{-3})\theta /\varepsilon$, providing an energy scale for which this state may be observed. In Fig.~\ref{fig1}, we plot the interaction potential given by Eq.~(\ref{potencial4}) for different values of $\theta$.
Note that the presence of the mass $\theta$ modifies the Coulomb potential by producing a region with $V(r)<0$, where bounded states are expected to emerge. The potential of Eq. (\ref{potencial4}) represents a competition between the repulsive Coulomb potential ($\propto 1/r$) and an attractive potential ($\propto\theta$). The final result presents a similar shape to the well known Lennard-Jones potential \cite{jbadams, ttsuneto}, which, in some cases, the attractive character is due to the electron-phonon interaction, just as the repulsive character is due to the electron-electron interaction.

It is interesting to note that, as highlighted in Ref.s~\cite{Yukawa,nascimento2015chiral}, if the mass of the PQED field is produced by an action that breaks the gauge invariance, as the Proca action $\sim M A_{\mu}A^{\mu}$, hence, the gauge-field propagator reads
\begin{equation}
\Delta^{(0)}_{\mu\nu}=\frac{\delta_{\mu\nu}}{2\sqrt{p^{2}}+M},
\end{equation}
where $M$ is the mass term. Therefore, the electron-electron interaction static potential, in this case with the breaking of gauge invariance, is given by the combination of the Coulombian potential with the Keldysh potential, namely,
\begin{equation}
V(r)=\frac{e^2}{4\pi}\left\lbrace\frac{1}{r}-\frac{\pi M}{4}\left[H_{0}\left(\frac{Mr}{2}\right)-Y_{0}\left(\frac{Mr}{2}\right)\right]\right\rbrace,
\end{equation}
where $H_{0}$ is a Struve function and $Y_{0}$ is a Bessel function. This electron-electron potential also has a competition between the Coulomb term and those proportional to $M$, nevertheless, it clearly does not generates bound states of particles with the same charge for any $M$ value.

Next, considering the Lagrangian in Eq.~(\ref{modelnm}), we investigate the effect of the vacuum polarization tensor on the photon propagator, and on the interaction potential between two charged particles.

The corrected gauge-field propagator is obtained from the Schwinger-Dyson equation
\begin{equation}\label{SDEQ}
\Delta_{\mu\nu}=\Delta_{\mu\alpha}^{(0)}(\delta_{\nu}^{\alpha}-\Pi^{\alpha\beta}\Delta_{\beta\nu}^{(0)})^{-1},
\end{equation}
where $\Delta_{\mu\alpha}^{(0)}$ is the free gauge-field propagator given by Eq.~(\ref{propagadordofoton}).

For electrons in the honeycomb lattice, we may use the two-component representation for the spinor, i.e, $\psi^{\dagger}_{a}=(\psi^{*}_{A}\psi^{*}_{B})_{a}$, where $(A, B)$ are the sublattices and $a=K,K^{\prime},\uparrow,\downarrow$ are the valley and spins indexes, respectively \cite{marino, PRX2015, review}. Therefore, we have $N_{f}=4$, which specifies to which valley the electron belong, as well as its spin orientation.

Substituting Eq.~(\ref{propagadordofoton}) in the Landau gauge and Eq.~(\ref{oneloop}) into Eq.~(\ref{SDEQ}), we find
\begin{equation}\label{EQTL}
\Delta_{\mu\nu}=\textrm{T}P_{\mu\nu}+\textrm{L}\epsilon_{\mu\nu\rho}p^{\rho},
\end{equation}
where
\begin{equation}
\textrm{T}=\frac{p^{2}}{2\varepsilon\sqrt{p^{2}}(p^{2}+\theta^{2})}\left\lbrace 1+b-c\theta\right\rbrace,
\end{equation}
and
\begin{equation}
\textrm{L}=\frac{1}{2\varepsilon\sqrt{p^{2}}(p^{2}+\theta^{2})}\left\lbrace \theta(1+b)+cp^{2}\right\rbrace.
\end{equation}
The auxiliary variables $b$ and $c$ are given by
\begin{equation}
b=-\frac{\Pi _2^2 p^2+2 \Pi _2 \theta  \sqrt{p^2}-2 \Pi _1 \sqrt{p^2}+\Pi _1^2}{4 \theta ^2+\left(\Pi _2^2+4\right) p^2-4 \left(\Pi _1-\Pi _2 \theta \right) \sqrt{p^2}+\Pi _1^2}
\end{equation}
and
\begin{equation}
c=\frac{2 \left(\Pi _1 \theta +\Pi _2 p^2\right)}{\sqrt{p^2} \left(4 \theta ^2+\left(\Pi _2^2+4\right) p^2-4 \left(\Pi _1-\Pi _2 \theta \right) \sqrt{p^2}+\Pi _1^2\right)}.
\end{equation}

Here, due to the spin and valley degeneracy, we must multiply Eq.s~(\ref{pi1}) and (\ref{pi2}) by $N_f=4$ in order to obtain $\Pi_{1}$ and $\Pi_{2}$. For simplicity, in this section, we will restrict our discussion to the case $m^{2}\ll p^{2}$. In this situation, up to first order in $m/p$, we have $\Pi_{1}=-e^{2}\sqrt{p^2}/4$ and $\Pi_{2}=e^{2}m/\sqrt{p^2}$. Therefore, Eq.~(\ref{EQTL}) reads
\begin{widetext}
\begin{equation}\label{rpaprop}
\begin{split}
\Delta_{\mu\nu}&=\frac{4 e^2 \sqrt{p^{2}} \left(\theta  (8 m-\theta )+p^2\right)+4 \sqrt{p^{2}} \left(8 \theta ^2+8 p^2\right)P_{\mu\nu}}{\epsilon  \left(\theta ^2+p^2\right) \left(e^4 \left(16 m^2+p^2\right)+16 e^2 \left(4 \theta  m+p^2\right)+64 \left(\theta ^2+p^2\right)\right)}+\\
&\frac{16 \left(e^2 m+2 \theta \right)\epsilon_{\mu\nu\alpha}p^\alpha}{\sqrt{p^{2}} \epsilon  \left(e^4 \left(16 m^2+p^2\right)+16 e^2 \left(4 \theta  m+p^2\right)+64 \left(\theta ^2+p^2\right)\right)}.
\end{split}
\end{equation}
\end{widetext}

Using an expansion, up to second order, for small amounts of $e$, we find
\begin{equation}
\begin{split}
\Delta_{\mu\nu}&=\frac{1}{2 \sqrt{p^2}\varepsilon \left(\theta ^2+p^2\right)}\left\lbrace \frac{p^{2}P_{\mu\nu}}{\left(1+\frac{e^2}{8}\right)}+\right.\\
&\left.\left[\theta-\frac{e^2 \left(2 \theta ^2 m-2 m p^2+\theta  p^2\right)}{4\left(\theta ^2+p^2\right)}\right]\epsilon_{\mu\nu\alpha}p^\alpha\right\rbrace. \label{DeltaF}
\end{split}
\end{equation}
Using Eq.~(\ref{DeltaF}), we may calculate the static potential, similarly to the calculations in the previous section, given by
\begin{equation}
V(r)=\dfrac{e^{2}}{4\pi\varepsilon\left(1+\frac{e^2}{8}\right)}\left\lbrace\dfrac{1}{r}+\frac{\theta\pi}{2}\left[ L_{0}\left(\theta r\right)-I_{0}\left(\theta r\right)\right]\right\rbrace.
\end{equation}
Note that the dielectric constant is modified due to the vacuum polarization effect, hence, providing an effective dielectric constant given by $\varepsilon_{eff}=\varepsilon(1+e^{2}/8)$, as expected. As an extra case, in appendix B, we calculate the gauge-field propagator in the 4x4 representation of Dirac matrices, using the RPA approach, and its respective static potential. Next, we calculate the electron self-energy of Eq.~(\ref{modelnm}).

\section{THE ANISOTROPIC ELECTRON SELF-ENERGY}\label{THE ANISOTROPIC ELECTRON SELF-ENERGY}

Here, we shall consider Eq.~(\ref{modelnm}) with a Lorentz symmetry breaking, which describes electrons that propagates with the Fermi velocity $v_F$ instead of the light velocity. This is easily performed by taking ${\partial}_{\mu}\rightarrow (\partial_{0},v_{F}\partial_{i})$ in the electron propagator and in the vertex $\gamma_{i}\rightarrow v_{F}\gamma_{i}$, for $c=1$,
\begin{equation}\label{vertice}
	\Gamma^{\mu}=e\left(\gamma^{0},v_{F}\gamma^{i}\right).
\end{equation}
and the fermion propagator Eq.~(\ref{fermionprop}),
\begin{equation}\label{anisofermionprop}
	S^{(0)}_{F}=\frac{-1}{\gamma^0 p_0+v_{F}\gamma^i p_i-m}.
\end{equation}

In this case, the static regime the electron self-energy $\Sigma$ is written as
\begin{equation}\label{autoenergia1}
\begin{split}
 \Sigma(p,m,\theta)&=\dfrac{Ce^{2}}{2\varepsilon}\int\dfrac{d^{D}k}{(2\pi)^{3}}\dfrac{1}{(\textbf{k}^{2})^{\frac{1}{2}}}\dfrac{\textbf{k}^{2}}{(\textbf{k}^{2}+\theta^{2})}\cdot \\
& \cdot\dfrac{(-\gamma^{0}(p_{0}-k_{0})+v_{F}\gamma^{i}(p_{i}-k_{i})-m)}{((p_{0}-k_{0})^{2}+v_{F}^{2}(\textbf{p}-\textbf{k})^{2}+m^{2})},
\end{split}
\end{equation}
where $D=d+1$ is the space-time dimension, and $C=\left(1+\frac{e^2}{8}\right)^{-1}$.

The denominator of Eq.~(\ref{autoenergia1}) has three terms. Therefore, we use the following Feynman parameterization
\begin{equation}\label{parametrizacao}
\dfrac{1}{a_{1}a_{2}a_{3}^{\frac{1}{2}}}=\frac{3}{4}\int_{0}^{1}dx\int_{0}^{1-x}dy\dfrac{(1-x-y)^{-\frac{1}{2}}}{[a_{1}x+a_{2}y+a_{3}(1-x-y)]^{\frac{5}{2}}}.
\end{equation}
The values of $a_{1}$, $a_{2}$ and $a_{3}$ are chosen as
\begin{equation}\label{a,b,c}
\begin{split}
a_{1} &= (p_{0}-k_{0})^{2}+v_{F}^{2}(\textbf{p}-\textbf{k})^{2}+m^{2},\\ 
a_{2} &= \textbf{k}^{2}+\theta^{2},\\ 
a_{3} &= \textbf{k}^{2}.
\end{split}
\end{equation} 
Making the shift $k_{0}\longrightarrow k_{0}+p_{0}$ in Eq.~(\ref{autoenergia1}) and eliminating the odd terms in $k_{0}$, due to the symmetric range of the integral, we have
\begin{equation}\label{autoenergia2}
\begin{split}
\Sigma(p,m,\theta) &= \frac{3Ce^{2}}{8}\int\dfrac{d^{D-1}k}{(2\pi)^{2}}\int_{0}^{1}dx\int_{0}^{1-x}dy\dfrac{(1-x-y)^{-\frac{1}{2}}}{x^{\frac{5}{2}}}\times \\
& \times\left\lbrace \vert\vec{k}\vert^{2}(v_{F}\gamma^{i}p_{i}-m)-\vert\vec{k}\vert^{2}v_{F}\gamma^{i}k_{i}\right\rbrace\times\\
&\times\int\dfrac{dk_{0}}{(2\pi)}\frac{1}{(k_{0}^{2}+\delta_{\vec{k}})^{\frac{5}{2}}},
\end{split}
\end{equation}
where 
\begin{equation}
\int\dfrac{dk_{0}}{(2\pi)}\frac{1}{(k_{0}^{2}+\delta_{\vec{k}})^{\frac{5}{2}}}=\frac{2}{3\pi\delta_{\vec{k}}^{2}},
\end{equation}
and the auxiliary functions $\delta_{\vec{k}}$, $\alpha_{x}$, and $\beta_{x}$ are defined as
\begin{equation}\label{delta,alphax,betax}
\begin{split}
\delta_{\vec{k}} &= \beta_{x}\left(k_{i}-\dfrac{xv_{F}^{2}}{\alpha_{x}}p_{i}\right)^{2}+v_{F}^{2}\left(1-\dfrac{xv_{F}^{2}}{\alpha_{x}}\right)+m^{2}+\dfrac{y}{x}\theta^{2}, \\ 
\alpha_{x} &= xv_{F}^{2}+(1-x), \\
\beta_{x} &= \dfrac{\alpha_{x}}{x}.
\end{split}
\end{equation}
After solving the integration over $k_{0}$ and performing the variable transformation $k_{i}\longrightarrow k_{i}-\dfrac{xv_{F}^{2}}{\alpha_{x}}p_{i}$ in Eq.~(\ref{autoenergia2}), we obtain
\begin{equation}\label{autoenergia3}
\begin{split}
&\Sigma(p,m,\theta) = \frac{Ce^{2}\mu^{\epsilon}}{4\pi}\int_{0}^{1}dx\int_{0}^{1-x}dy\dfrac{(1-x-y)^{-\frac{1}{2}}}{x^{\frac{5}{2}}\beta_{x}^{2}}\times \\
& \times\left\lbrace(v_{F}\gamma^{i}p_{i}-m-\dfrac{2xv_{F}^{3}}{\alpha_{x}}\gamma^{i}p_{i})\int\dfrac{d^{D-1}k}{(2\pi)^{2}}\dfrac{k_{i}^{2}}{(k_{i}^{2}+\Delta)^{2}}+\right. \\ 
& \left. +\dfrac{x^{2}v_{F}^{4}p_{i}^{2}}{\alpha_{x}^{2}}(v_{F}\gamma^{i}p_{i}-m+\dfrac{xv_{F}^{3}}{\alpha_{x}}\gamma^{i}p_{i})\int\dfrac{d^{D-1}k}{(2\pi)^{2}}\frac{1}{(k_{i}^{2}+\Delta)^{2}}\right\rbrace ,
\end{split}
\end{equation}
where $\Delta=[v_{F}^{2}(1-\frac{xv_{F}^{2}}{\alpha_{x}})p_{i}^{2}+m^{2}+\frac{y}{x}\theta^{2}]/\beta_{x}$. Note that, we will use the dimensional regularization scheme. Hence, we made $e\rightarrow \mu^{\epsilon /2}\,e$, where $\mu$ is the scale parameter and $\epsilon$ is the dimensional regulator, such that $d=2+\epsilon$. In this case, the first integral in $d^{2}k$ presents a logarithmic divergence
\begin{equation}
\int\dfrac{d^{D-1}k}{(2\pi)^{2}}\dfrac{k_{i}^{2}}{(k_{i}^{2}+\Delta)^{2}}=\frac{\Gamma(\frac{\epsilon}{2})}{(4\pi)^{(2+\epsilon)/2}}.
\end{equation}
On the other hand, the second one is finite, given by
\begin{equation}
\int\dfrac{d^{D-1}k}{(2\pi)^{2}}\dfrac{1}{(k_{i}^{2}+\Delta)^{2}}=\frac{1}{4\pi\Delta}.
\end{equation}

Next, solving the integrals over the Feynman parameters ($y$, $x$), we have
\begin{equation}\label{autoenergia4}
\Sigma(p,m,\theta)=-\dfrac{Ce^{2}}{4\pi}\left\lbrace\dfrac{1}{4}\gamma^{i}p_{i}+\dfrac{m}{2v_{F}}\right\rbrace\dfrac{1}{\epsilon}+\textrm{finite terms}.
\end{equation}
Note that, due to the $k_{i}^{2}$ factor, from the gauge-field propagator in Eq.~(\ref{propagadordofoton}), the dimensional integral over the $k_{i}^{2}$-term eliminates the dependence on the $\theta$ parameter in the divergent part of the self-energy. It is also important to note that the dependence of the topological mass $\theta$ is restricted to the finite terms, hence, the renormalized quantities like Fermi velocity $v_{F}^{R}$ and mass $m^{R}$ are not dependent on $\theta$ in the light of the renormalization group equations.

Using the renormalization group method by following Ref.~\cite{pqed+cs}, we obtain that
\begin{equation}\label{betavf}
\beta_{v_{\textrm{F}}}=\mu\frac{\partial v_{F}}{\partial\mu}=-\frac{e^2}{16\pi}
\end{equation}
and
\begin{equation}\label{betam}
\beta_{m}=\mu\frac{\partial m}{\partial\mu}=-\frac{me^2}{8\pi v_{F}}.
\end{equation}
For more details about the renormalization group calculations, see Appendix A.

Using Eq.~(\ref{betavf}), the flow of the effective Fermi velocity may be written as
\begin{equation}\label{vfrenor}
v_{F}^{R}(\mu) = v_{F}(\mu_{0})\left[1-\dfrac{\alpha}{4}\ln\left(\frac{\mu}{\mu_0}\right)\right]
\end{equation}
and, using Eq.~(\ref{betam}), the mass is
\begin{equation}\label{mrenor}
m^{R}(\mu) = m(\mu_0)\left(\frac{\mu}{\mu_0}\right)^{-\alpha/2},
\end{equation}
where $\alpha$ is the fine structure constant, namely, $\alpha=e^{2}/4\pi \varepsilon_{eff}v_{F}$, such that the effective dieletric constant reads $\varepsilon_{eff}=\varepsilon\left(1+e^{2}/8\right)$.

The expression of the renormalized Fermi velocity is well known in the literature and it has been experimentally observed in suspended graphene \cite{vozmediano, Gorbachev}. More recently, it has been shown that electromagnetic interactions also provides an useful framework to explain the mass renormalization, which has been experimentally observed in a few TMDs, as discussed in Ref.~\cite{luis}. In this case, the authors have considered the large-$N$ expansion and find a good agreement with the experimental data. The main feature of $m_R$ is that it decreases as we increase the energy scale $\mu$. Here, considering our result in Eq.~(\ref{mrenor}), we may conclude that this behavior is preserved regardless of the presence of $\theta$ and even within the perturbation theory, whose result could be schematically obtained only by taking $\varepsilon_{eff}\rightarrow\varepsilon$ in Eq.~(\ref{mrenor}), i.e, by neglecting the term generated by $\Pi_1$. A similar result also holds for $v^R_F$. Obviously, the reason for such invariance is connected to the divergent terms that are relevant for the renormalization group equations.

\section{DISCUSSION}\label{DISCUSSION}

In this work, we study some aspects of a generalization of the Chern-Simons action, containing the pseudo-differential operator $\sqrt{-\Box}$. From the bosonization of free massive Dirac particles in (2+1)D, we show that, in the limit when $m^{2}\ll p^{2}$, the bosonized theory is equivalent to PQED plus a PCS, given by Eq.~(\ref{Zh}). On the other hand, in the limit $m^{2}\gg p^{2}$, we obtain a Maxwell-Chern-Simons theory, given by Eq.~(\ref{R11}), where $m$ is the fermion mass and $p$ is its momentum. This generalizes the result obtained for massless fermions in Ref.~\cite{marino}. Otherwise, Ref.~\cite{PLBOzela} finds the same theory as a dual-action to the Chern-Simons-Higgs theory. However, this procedure used by the authors did not cause any limitation to the theory massive regime, nor the Dirac matrices representation.

We also investigated the role of the PCS when combined to PQED theory. At tree level, the static potential between the electrons, in Eq.~(\ref{potencial4}), is short-range due to the mass in the mediating field. Furthermore, it presents a confinement region between an electron pair, given by a non-symmetric potential well around the stable equilibrium point. This is generated by the competition between the repulsive Coulomb potential of the PQED and the attractive potential associated with the PCS theory, see Fig.~\ref{fig1}. These electron-electron bonded states have a minimum binding energy in order of $-3.0\times 10^{-3}\theta$. From the quantum corrections to the static potential, we conclude that the insertion of the vacuum polarization tensor in the mediating field produces an effective dielectric constant, given by $\varepsilon_{eff}=\varepsilon (1+e^{2}/8)$. Therefore, the quantum fluctuations yields a suppression of the static potential, similarly to the standard screening effects in the Coulomb potential. Our static potential $V(r)$ is also expected to provide new physical results when considering the interaction of a point-like Dirac particle and a conducting surface as discussed in Ref.~\cite{borges2020new} for the case of the Maxwell-Chern-Simons theory.

As a final aspect of our model, we calculate the electron self-energy at one-loop perturbation theory, using the dimensional regularization scheme. Thereafter, we consider the renormalization group equation for the renormalized vertex function, from which we show that the PCS term does not change the renormalization of both the Fermi velocity and of the band gap in a Dirac-like material.

Despite the presence of a pseudo-differential operator in the PCS action, this term provides several interesting features when coupled to PQED. In particular, it does not change some results that have been already confirmed by some independent experimental data. Furthermore, it includes a mass term for the gauge-field propagator, in analogy to the topological mass of the Maxwell field, and predicts the realization of bounded particles with the same charge. A simple application of this result would be to consider the generation of Cooper pairs for describing a superconductor phase in the honeycomb lattice. This may be relevant for describing non-BCS superconductors. We shall investigate this elsewhere.

\section*{ACKNOWLEDGMENTS}

G. C. M.  is partially supported by Coordenação de Aperfeiçoamento de Pessoal de Nível Superior – Brasil (CAPES), finance code 001; V. S. A. and L. O. N. are partially supported by Conselho Nacional de Desenvolvimento Científico e Tecnológico (CNPq) and by CAPES/NUFFIC, finance code 0112; E. C. M. is partially supported by both CNPq and Fundação Carlos Chagas Filho de Amparo à Pesquisa do Estado do Rio de Janeiro (FAPERJ). The authors also thank R. F. Ozela for his useful comments.

\section*{APPENDIX A: RENORMALIZATION GROUP CALCULATIONS}

We start with The 't Hooft-Weinberg renormalization group equation, given by
\begin{equation}\label{eqgr}
\left(\mu\frac{\partial}{\partial\mu}+\beta_{v_{\textrm{F}}}\frac{\partial}{\partial v_{F}}+\beta_{m}\frac{\partial}{\partial m}-N_F\gamma_{\psi}\right)\Gamma_{\textrm{R}}^{(N_F,N_A)}=0,
\end{equation}
where $\Gamma_{\textrm{R}}^{(N_F,N_A)}$ is the renormalized vertex function with $N_F=2$ fermion fields and $N_A=0$ gauge fields. Note that the electric charge and the wavefunction of the gauge field are not renormalized.  The renormalization group functions are usually defined as $\beta_{v_{\textrm{F}}}=\mu \frac{\partial v_{\textrm{F}}}{\partial\mu}$, $\beta_{m}=\mu \frac{\partial m}{\partial\mu}$, and $\gamma_{\psi}$ is the anomalous dimension of the matter field. For calculating $\Gamma_{\textrm{R}}^{(2,0)}$, we use the prescription 
\begin{equation}
(1-{\cal T})\mu^{\epsilon}\,I^{(2,0)}=\mbox{Finite}^{(2,0)}+\ln\mu\,\mbox{Res}^{(2,0)}
\end{equation}
in order to remove the divergent term in the electron self-energy $I^{(2,0)}$, where ${\cal T}$ is the Taylor operator that removes the pole. On the other hand, the factor $\mbox{Res}^{(2,0)}$ is the residue of the diagram given by the coefficient of $1/\epsilon$, as well as $\mbox{Finite}^{(N_F,N_A)}$ is the finite part of the amplitude $I^{(2,0)}$. Using this prescription we find that, for one-loop calculation,
\begin{equation}\label{gammar}
\Gamma_{\textrm{R}}^{(2,0)}(\bar{p})= -(\gamma^{\mu}\bar{p}_{\mu}-m) +\Sigma (\bar{p}) \ ,
\end{equation}
where $\Sigma (\bar{p}) = e^2\left[\mbox{Finite}^{(2,0)}+\ln\mu\,\mbox{Res}^{(2,0)}\right]$, and $\mbox{Res}^{(2,0)}=A_1\,\gamma^0p_0+A_2\,\gamma^i p_i+A_3$. 

Because we consider an one-loop expansion, we can write $\beta_{v_{\textrm{F}}}=e^{2}\beta_{v_{\textrm{F}}}^{(2)}$, $\beta_{m}=e^{2}\beta_{m}^{(2)}$, and $\gamma_{\psi}=e^{2}\gamma_{\psi}^{(2)}$. Thereafter, using Eq.~(\ref{gammar}) in Eq.~(\ref{eqgr}), up to order $e^2$, we have $
\gamma_{\psi}^{(2)}=\frac{1}{2}A_1$, $\beta_{v_{\textrm{F}}}^{(2)}= A_2-v_{\textrm{F}}\,A_1$, and $\beta_{m}^{(2)}= A_1+A_3$. From the electron self-energy in Eq.~(\ref{autoenergia4}), we obtain that $A_1= 0$, $A_2= -\frac{1}{16\pi}$, $A_3 = -\frac{m}{8\pi v_{F}}$. Therefore, it follows that Eq.(\ref{betavf}) and Eq.(\ref{betam}).

\section*{APPENDIX B: SCREENING EFFECT FOR THE 4x4 DIRAC MATRICES AND RPA APPROACH}\label{4x4}

In this appendix, we investigate the effects of the vacuum fluctuations on the gauge-field propagator, using the random phase approximation (RPA), and hence on the interaction potential between two charged particles.

\begin{figure}[h]
\centering
\includegraphics[scale=.66]{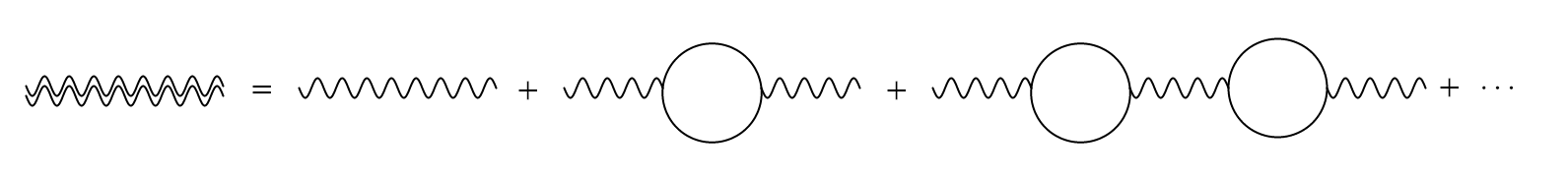}
\caption{Illustration of the RPA method. The gauge field propagator corrected by the vacuum polarization tensor. In this illustration, the bubbles represent the vacuum polarization tensor.}\label{rpasum}
\end{figure}

In the RPA method, we incorporate the effects of electron-electron interaction in the free propagator of the gauge field through the sum of infinite diagrams, corrected by the polarization tensor. Fig.~\ref{rpasum} show this sum of the leading-order terms.

This sum results in Eq.~(\ref{SDEQ}). Therefore, it also results in Eq.~(\ref{EQTL}). Nevertheless, in the four-component representation for the Dirac matrices, $\Pi_{2}=0$, and the polarization tensor is written as
\begin{equation}\label{PolarizationTensor}
\Pi^{\mu\nu}(p,m)=\Pi_{1}(p,m)P^{\mu\nu},
\end{equation}
where
\begin{equation}\label{Pi}
\Pi_{1}(p,m)=-\frac{2e^{2}}{\pi}p^{2}\int_{0}^{1}dx\frac{x(1-x)}{\sqrt{x(1-x)p^{2}+m^{2}}},
\end{equation}
where we have use that $N_{f}=2$. In this case, we find
\begin{equation}
\textrm{T}=\frac{8(p^{2})^{3/2}+\theta^{2}(8\sqrt{p^{2}}-4\Pi_{1})+4p^{2}\Pi_{1}-\Pi_{1}^{2}(2\sqrt{p^{2}}+\Pi_{1})}{16(p^{2}+\theta^{2})^{2}+8(\theta^{2}-p^{2})\Pi_{1}^{2}+\Pi_{1}^{4}}
\end{equation}
and
\begin{equation}
\textrm{L}=\frac{2\theta}{4((p^{2})^{3/2}-p^{2}\Pi_{1})+\sqrt{p^{2}}(4\theta^{2}+\Pi_{1}^{2})}.
\end{equation}

For simplicity, we will consider only the case of massless fermions. Hence, $\Pi_{1}(p,m=0)=-e^2\sqrt{p^{2}}/4$. Therefore, Eq.~(\ref{EQTL}) reads
\begin{equation}\label{rpaprop}
\Delta_{\mu\nu}=\frac{p^{2}P_{\mu\nu}+\frac{\theta}{(1+\frac{e^{2}}{8})}\epsilon_{\mu\nu\rho}p^{\rho}}{2\varepsilon_{eff}\sqrt{p^{2}}\left[p^{2}+\frac{\theta^{2}}{(1+\frac{e^{2}}{8})^{2}}\right]}.
\end{equation}
Note that due to the screening effect, the massive pole is scaled by $\theta\rightarrow\bar{\theta}=\varepsilon\theta/\varepsilon_{eff}$, where $\varepsilon_{eff}=\varepsilon (1+e^{2}/8)$ may be understood as an effective dielectric constant. Note that for $\theta=0$ Eq.~(\ref{rpaprop}) reproduces the gauge-field propagator of PQED in the RPA \cite{CSBPQED}.

From Eq.~(\ref{rpaprop}), we obtain the static potential with quantum corrections as it has been obtained in \cite{PLBOzela}. Due to the form of the corrected propagator, the result has the same form as in Eq.~(\ref{potencial4}), in which we associate $\theta\rightarrow\bar{\theta}$ and the appearance of an overall factor, $(1+e^{2}/8)^{-1}$. Having this in mind, it is straightforward that the corrected static potential reads
\begin{widetext}
\begin{equation}\label{potencialrpa}
V(r)=\dfrac{e^{2}}{4\pi\varepsilon_{eff}}\left\lbrace\dfrac{1}{r}+\frac{\theta\pi}{2(1+\frac{e^{2}}{8})}\left[ L_{0}\left(\frac{\theta r}{1+\frac{e^{2}}{8}}\right)-I_{0}\left(\frac{\theta r}{1+\frac{e^{2}}{8}}\right)\right]\right\rbrace.
\end{equation}
\end{widetext}
From the corrected static potential, we conclude that the potential well depth, which is proportional to the massive pole of the gauge-field propagator, will be inversely proportional to the factor $\left(1+e^{2}/8\right)^{2}$, hence, reducing the number of bound states on the well.

\bibliography{refs}

\clearpage

\end{document}